\documentclass{article}

\usepackage[final]{neurips_2020_ml4ps}

\usepackage[utf8]{inputenc} 
\usepackage[T1]{fontenc}    
\usepackage{url}            
\usepackage{booktabs}       
\usepackage{amsfonts}       
\usepackage{nicefrac}       
\usepackage{microtype}      

\usepackage{amsmath}

\usepackage{graphicx}
\usepackage{subcaption}
\usepackage{caption}
\usepackage{xcolor}
\usepackage{wrapfig}
\usepackage[frozencache,cachedir=.]{minted}

\usepackage{color}
\usepackage{xcolor,colortbl}
\definecolor{matlab-blue}{rgb}{0,0.4470,0.7410}
\definecolor{matlab-orange}{rgb}{0.8500,0.3250,0.0980}
\definecolor{matlab-yellow}{rgb}{0.9290,0.6940,0.1250}
\definecolor{matlab-green}{rgb}{0.4660,0.6740,0.1880}
\definecolor{matlab-red}{rgb}{0.6350,0.0780,0.1840}
\definecolor{matlab-purple}{rgb}{0.4901,0.1803,0.5529}
\definecolor{ourmethod}{gray}{0.93}
\definecolor{mydarkblue}{rgb}{0,0.08,0.45}

\newcommand{\etal}{\textit{et al.}}

\definecolor{light-gray}{gray}{0.95}
\newcommand{\code}[1]{\colorbox{light-gray}{\texttt{#1}}}
\newcommand{\AMPERE}{\code{AMPERE}~}
\newcommand{\SUPERMAG}{\code{SuperMAG}~}
\newcommand{\MHD}{\code{MHD}~}
\newcommand{\OMNI}{\code{OMNI}~}

\newcommand{\WIEMER}{\cite{weimer2013empirical} model~}

\newcommand{\SUPERMAGIMRPOVEMENT}{{14.53\%}}
\newcommand{\MHDIMRPOVEMENT}{{24.35\%}}

\newcommand*\samethanks[1][\value{footnote}]{\footnotemark[#1]}

\title{Global Earth Magnetic Field Modeling and Forecasting with Spherical Harmonics Decomposition}

\author{%
  Panagiotis Tigas\thanks{equal contribution} \\
  OATML \\
  University of Oxford\\
  Oxford, UK
  \And
  To Bloch\samethanks \\
  University of Reading \\
  Reading, UK
  \And
  Vishal Upendran\samethanks \\
  IUCAA \\
  Pune, India
  \And
  Banafsheh Ferdoushi\samethanks \\
  University of New Hampshire \\
  Durham, NH, USA
  \And
  Yarin Gal \\
  OATML \\
  University of Oxford \\
  Oxford, UK
  \And
  Siddha Ganju \\
  NVIDIA Corporation \\
  Santa Clara, CA, USA
  \And
  Ryan M. McGranaghan \\
  ASTRA LLC \\
  Louisville, CO, USA
  \And
  Mark C. M. Cheung \\
  Lockheed Martin\\Advanced Technology Center\\
  Palo Alto, CA, USA
  \And
  Asti Bhatt \\
  SRI International \\
  Menlo Park, CA, USA
}

\begin{document}

\maketitle

\begin{abstract}
  Modeling and forecasting the solar wind-driven global magnetic field perturbations is an open challenge. Current approaches depend on simulations of computationally demanding models like the Magnetohydrodynamics (\MHD) model or sampling spatially and temporally through sparse ground-based stations (\SUPERMAG). In this paper, we develop a Deep Learning model that forecasts in Spherical Harmonics space \footnote{We will release all the code and models used in this work post-publication.}, replacing reliance on \MHD models and providing global coverage at one-minute cadence, improving over the current state-of-the-art which relies on feature engineering. We evaluate the performance in \SUPERMAG dataset (improved by \SUPERMAGIMRPOVEMENT) and \MHD simulations (improved by \MHDIMRPOVEMENT). Additionally, we evaluate the extrapolation performance of the spherical harmonics reconstruction based on sparse ground-based stations (\SUPERMAG), showing that spherical harmonics can reliably reconstruct the global magnetic field as evaluated on \MHD simulation.

\end{abstract}

\section{Introduction}

The space environment around Earth (geospace) does not exist in a steady state. The dynamical changes in geospace is termed space weather. Space weather is primarily driven by interactions between the Sun's output (in terms of solar radiation and the magnetized solar wind) with the Earth's magnetosphere, thermosphere and ionosphere. The US National Oceanic and Atmospheric Administration continuously monitors the solar wind upstream from Earth with satellite observatories orbiting Lagrangian point L1 along the Sun-Earth line. The solar wind energy is transferred to the Earth's magnetosphere via complex mechanisms~\citep{dungey1961interplanetary}, leading to perturbations in the Earth's magnetic field called geomagnetic storms.

Geomagnetic storms drive a spectrum of potentially catastrophic disruptions to our technologically-dependent society, among the most threatening being critical disturbances to the electrical grid in the form of \textit{geomagnetically induced currents} (\code{GICs}). Due to their proprietary nature, publicly available \code{GIC} data are limited. However, a cohort study of insurance claims of electrical equipment provides evidence that space weather poses a continuous threat to electrical distribution grids via geomagnetic storms and \code{GICs}~\citep{Schrijver:2014, eastwood2018quantifying}. \code{GICs} also pose threats to oil pipelines, railyways and telecommunication systems.
In the case of extreme, but historically probable geomagnetic storms, the economic impact due to prolonged power outages can exceed billions of dollars per day ~\citep{Oughton:2017}. For this reason, there is urgency among public and industry stakeholders to improve monitoring and forecasting of space weather impacts like geomagnetic storms and \code{GICs}.

The geomagnetic field is continuously monitored, importantly with sparse spatial coverage, by a network of roughly 200 ground magnetometers~\citep{gjerloevSuperMAGDataProcessing2012} and by a constellation of 66 satellites collectively known as the \textit{Active Magnetosphere and Planetary Electrodynamics Response Experiment} (\AMPERE) project ~\citep[AMPERE;][]{Waters:2020} in low Earth orbit. The spatially-sparse magnetometer measurements are typically synthesized into global indices (e.g. \textbf{Dst}, \textbf{Kp}, and \textbf{AE} ; see appendix for more details) as measures of the \emph{geoeffectiveness} of space weather perturbations. Like most indices, they are good as indicators but are too far removed from the underlying governing equations of the system. This makes interpretability and forecasting (whether by physics-based or physics-agnostic ML models) difficult.


GICs are driven by geoelectric fields through Ohms law, $\vec{J} = \mathbf{\sigma} \vec{E}$ (the displacement current is negligible), where $\vec{J}$ is the current, $\mathbf{\sigma}$ the conductivity tensor, and $\vec{E}$ is the electric field. The geoelectric field is related to the ground magnetic field perturbations through the Faradays law of induction $ - \nabla \times \vec{E} = \frac{\partial \vec{B}}{\partial t}$, where, $\frac{\partial \vec{B}}{\partial t}$ is the rate of change of magnetic field on the Earth's surface and can be derived from the time derivative of magnetic perturbation based on ground Earth magnetometers.

\begin{figure}[t!]
  \centering
  \hspace*{-1.7cm}
  \includegraphics[width=1.2\textwidth]{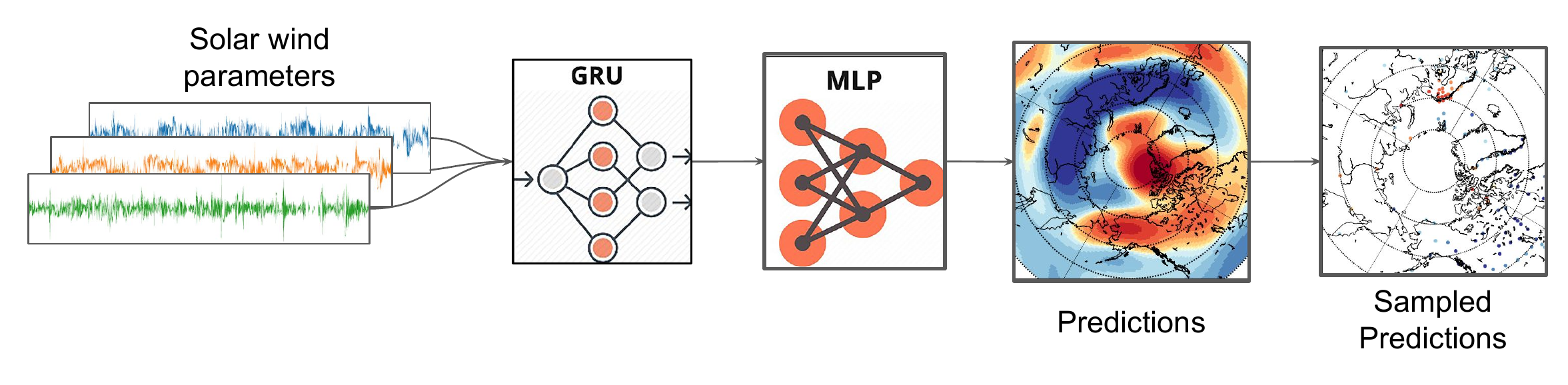}
  \caption{
    Model architecture. Solar Wind data are summarized into an embedding vector using a \textit{Gated Recurrent Unit} (GRU) which is then passed to a \textit{Multilayer Perceptron} (MLP) to output the predicted spherical harmonics coefficients from where we can extrapolate values for specific locations.
  }
  \label{fig:architecture}
  \vspace{-1.0em}
\end{figure}

\textbf{Our contributions are:}
\vspace{-0.5em}
\begin{enumerate}
    \item Using simulation data from a physics-based (\MHD) model to validate a compressed sensing technique to recover global maps (in spherical harmonic basis) of the geomagnetic perturbation from sparse measurements. This improves the temporal cadence of such maps by $>$10x. \vspace{-0.5em}
    \item Developing a Deep Learning model that operates in the spherical harmonics space, allowing for global modeling of the magnetic field disturbances, combined with powerful non-linear autoregressive models (RNN, 1D-CNN) to capture the influence of solar wind data.
\end{enumerate}
\vspace{-1.0em}

\section{Datasets}

We used data describing Earth's magnetosphere and solar wind properties between \texttt{1 Jan 2013} and \texttt{31 Dec 2013}. We used \OMNI\; dataset for solar wind data, which captures the interplanetary magnetic field components (IMF), velocity and temperature of the solar wind as well as the clock angle. \footnote{Solar wind measurement is performed by ACE and WIND satellites at the L1 Lagrange point of the Sun-Earth system}. For Earth's magnetic field, we used \SUPERMAG\; dataset~\citep{gjerloev2012supermag} which consists of geomagnetic perturbations measurements from ground earth stations located at various places around the globe.



Finally, we used a simulation-derived dataset based on simulations conducted with Open Geospace General Circulation Model (\code{Open GGCM})~\citep{raeder2001geomagnetic}, a magnetohydrodynamics (\MHD) model. Lacking a global and spatially-complete ground-based magnetometer array we must rely on the first principles \MHD\; model to constitute a global ground truth dataset. Furthermore, this allows us to validate our approach on finer resolution than one supported by \SUPERMAG\; and \AMPERE\; dataset. However, note that the \MHD\; model is not a substitute for actual perturbation measurements, for many short-scale phenomena are missed by such models~\citep{raeder2001geomagnetic}.



\section{Methodology}

In order to properly forecast GICs, we need to be able to model the relationship between the solar wind parameters and interplanetary magnetic fields (IMF) with the Earth system. Due to the scarcity of GIC data, the comparatively abundant magnetometer data and the relationship between the two, a logical step is to create models to predict $\frac{\partial \vec{B}}{\partial t}$. Such models provide complete spatial coverage and permit the study of the nature of the connection between the solar wind and the Earth.

\subsection{Spherical Harmonics Decomposition}

\begin{wrapfigure}{r}{0.23\textwidth}
    \centering
    \vspace{-2.7em}
    \includegraphics[scale=0.23]{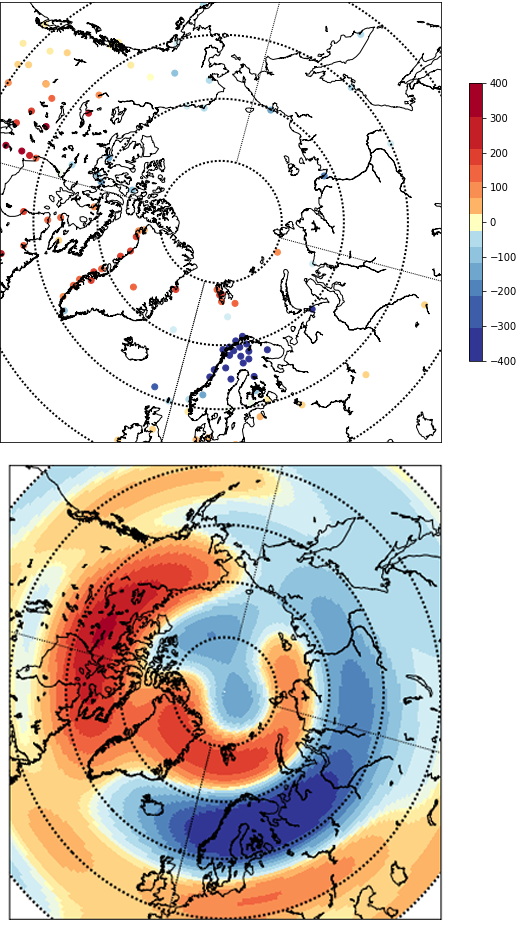}
    \caption{Ground truth observations (top) and Spherical harmonic reconstruction (bottom) of \texttt{dbn\_nez}.}
    \label{fig:supermag_recons}
    \vspace{-6.9em}
\end{wrapfigure}

Any scalar field over the unit sphere can be expressed as $f(\theta,\phi) = \sum_{n=0}^{\infty} \sum_{m=-n}^{n} a_{nm} Y_{nm}(\theta,\phi)$, where $$Y_{nm}(\theta,\phi) := \color{gray} \sqrt{\frac{2n+1}{4\pi} \frac{(n-m)!}{(n+m)!}} \color{black} e^{i m \theta} P^m_n(\cos(\phi)),$$ and $P^m_n(\cos(\phi))$ are the associated Legendre polynomials. These  functions $Y_{nm}(\theta,\phi)$ are solutions to Laplace equation in a 3-D spherically symmetric coordinate system. If the sum is truncated at maximum harmonic degree $N$, $f(\theta,\phi)$ is approximated as  $\tilde{f}(\theta,\phi) = \sum_{n=0}^{N} \sum_{m=-n}^{n} a_{nm} Y_{nm}(\theta,\phi)$. Defining $i = n^2 + n + m$, the double sum can be written as  $\tilde{f}(\theta,\phi) = \sum_{i=0}^{(N+1)^2-1} a_{i} Y_{i}(\theta,\phi)$. If the 2D fields over $\theta,\phi$ are unrolled as one-dimensional arrays, we have  $\tilde{f} = \mathcal{B} \vec{a}$, where $\vec{a}=(a_i)$ is the tuple of spherical harmonic coefficients, and  $\mathcal{B} = (\vec{b}_i)$ is the basis matrix wherein column vector $\vec{b}_i$ represents the basis function $Y_{nm}(\theta,\phi$).

\subsection{Experimental Design}

\subsubsection{Reconstruction}


With \SUPERMAG data, we have spatially sparse samples $y$ which can be any component of the geomagnetic field, or its deviation from a reference field) measured at \~200 stations. Note that only the Northern Hemisphere is considered in this analysis, for it has an extensive coverage (in contrast to the extremely sparse coverage in the Southern Hemisphere). The spherically symmetric reconstruction thus is $\tilde{f}(\theta,\phi)$, and this reconstruction as evaluated at the station locations is $\tilde{y}$. Thus, we first construct a set of coefficients $\vec{a}$ from this sparse set of measurements, considered all the measurements across the globe at each time step. There are two constraints imposed for obtaining the set of coefficients:

\begin{wrapfigure}{r}{0.6\textwidth}
    \centering
    \vspace{-1.7em}
    \includegraphics[width=0.48\linewidth]{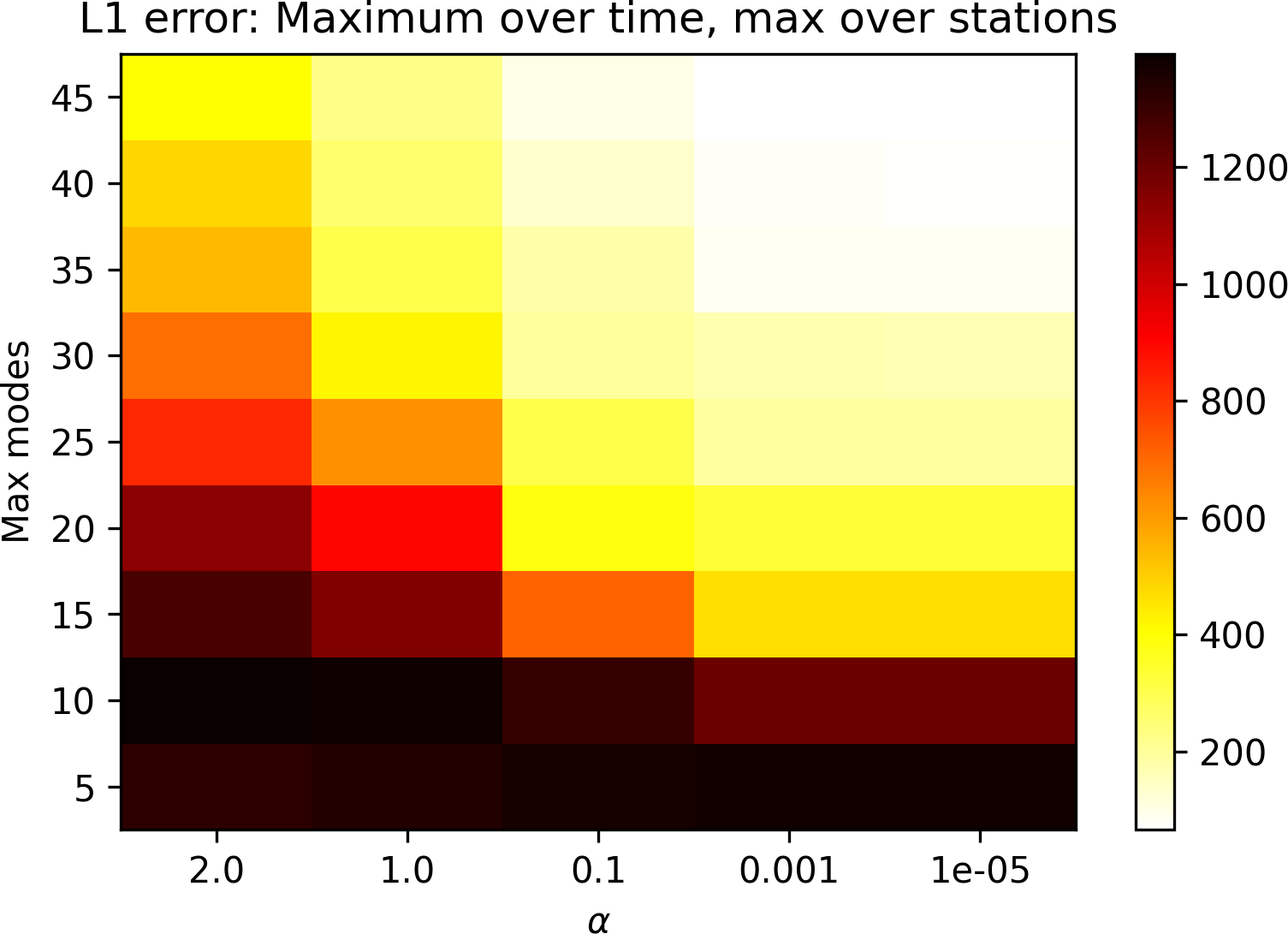}
    \includegraphics[width=0.48\linewidth]{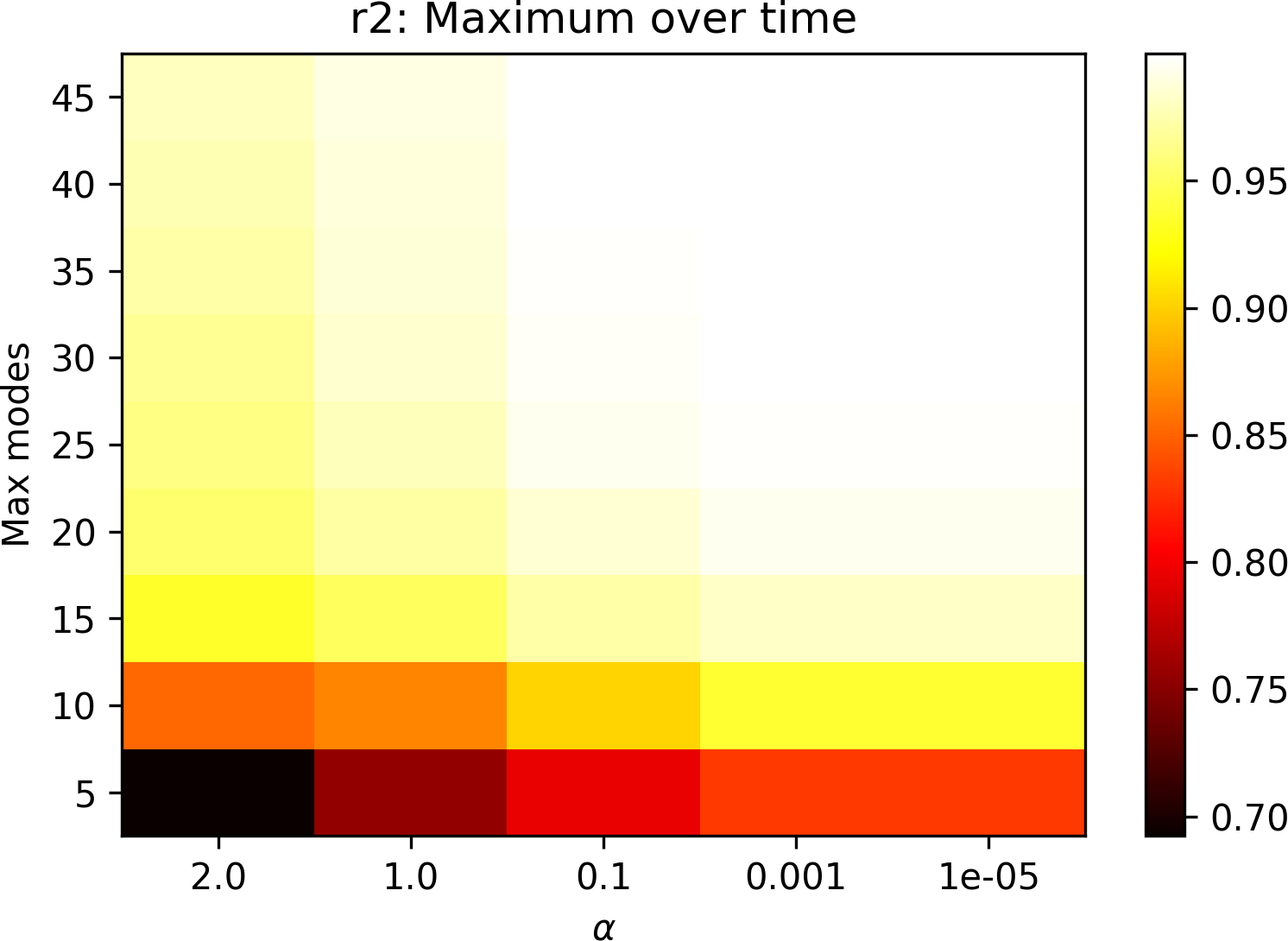}
    \caption{L1 error (left) and R2 (right) "knee" determination.}
    \label{fig:L1_R2_knee}
    \vspace{-0.4em}
\end{wrapfigure}

\vspace{-4pt}
\begin{enumerate}
    \item The coefficient set must be sparse: This is to prevent power leakage and coupling across multiple modes.
    \item As small $N$ as possible: Since, in theory, an infinite number of modes can fit the observations, but a parsimonious model is required to limit overfitting. Furthermore, we would not want unnatural, localized artifacts due to high number of modes, thereby motivating a constraint on $n$.
\end{enumerate}
\vspace{-4pt}
To mitigate constraint ($1$), we apply a Lasso regression technique~\citep{Lasso} with the spherical harmonic functions as the basis.

The Lasso regression comes with a regularization term $\alpha\lVert \vec{a}\rVert_1$, where $\lVert \vec{a}\rVert_1$ is the L1-norm of the coefficients and $\alpha>0$ is a hyperparameter. This, and constraint ($2$) are mitigated by varying both $\alpha$ and the maximum number of modes $N$, and searching for a knee in a defined error metric, subject to the smallest maximum modes. The sweep parameters are detailed in Table.~\ref{reconstruction_sweep}, and two metrics are used -- Maximum L1 error across all stations and time, and maximum R2 metric (coefficient of determination) across all time steps. The maximum L1 error tells us the worst performance across the dataset, and thus we seek the most acceptable worst possible performance. As shown in Fig.~\ref{fig:L1_R2_knee}, the "knee of goodness" can be seen corresponding to $\alpha = 0.1$, and 20 maximum modes. These parameters are fixed in our analysis.

\begin{figure}[!t]
    \centering
    \includegraphics[width=0.65\linewidth]{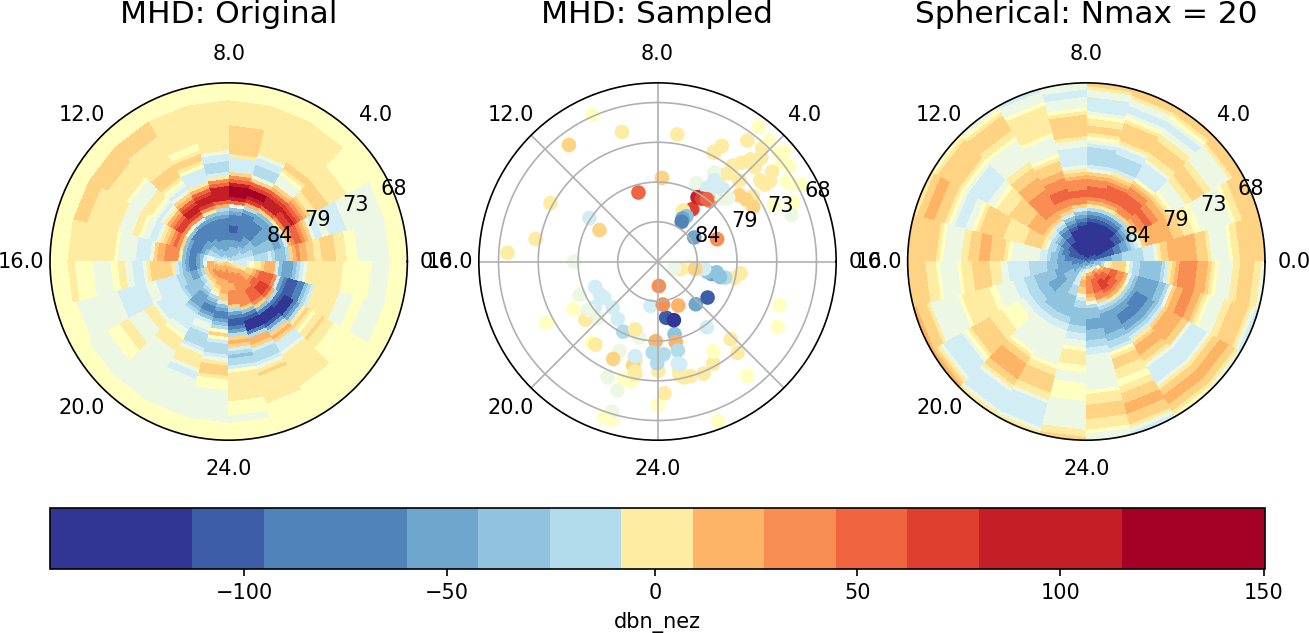}
    \caption{Comparison of reconstruction with the MHD simulation. Left: MHD simulation; center: MHD simulation sampled at SuperMAG station locations; right: Reconstruction.}
    \label{fig:mhdreconstr}
    \vspace{-1em}
\end{figure}

An example reconstruction of the data from 2017, considering stations above $40^\circ$ is shown in Fig.~\ref{fig:supermag_recons}. Here, we compare the north facing component of $\partial d / \partial t$ from \SUPERMAG and its reconstruction. Similarly, the comparison of spherical harmonic reconstruction for the \MHD simulation is shown in Fig.~\ref{fig:mhdreconstr}. Note here, that the reconstruction is performed by sampling the \MHD simulation at locations of \SUPERMAG stations alone.


\subsubsection{Forecasting}

Next, we construct a forecasting model which uses solar wind data (\OMNI) to forecast the global magnetic field perturbation. For this experiment we use a similar setup as \WIEMER. We feed 25 minutes of solar wind activity into a Gated Recurrent Network (GRU) to map the sequence into an embedding vector. We then feed the embedding into a Multilayer Perceptron (MLP) to output the spherical harmonics coefficients which model the global magnetic field perturbation; specifically we focus on north component of $\partial d / \partial t$ as a proof of concept. In contrast to \WIEMER, we use the whole sequence as input to a non-linear autoregressive model and we do not apply feature engineering to the \OMNI data. Instead we use the raw features (see appendix for detailed list of features). The architecture can be seen in fig.~\ref{fig:architecture}.


We benchmark this work against the state-of-the-art \emph{empirical} model by Wiemer \etal~\citep{weimer2013empirical}. To evaluate the performance we first compare on the validation set on \SUPERMAG but also we compare on simulations conducted with \MHD model for two weeks worth of activity. The results are summarized in table~\ref{performance}.



\section{Results And Discussion}

\vspace{-1em}
\begin{table}[!h]
  \caption{Forecasting model performance}
  \label{performance}
  \centering
  \begin{tabular}{lrr}
    \toprule
    Model       & \SUPERMAG (val) RMS (nT) $\downarrow$ & \MHD RMS (nT) $\downarrow$ \\
    \midrule
    Ours.       & \textbf{24.23}                                   & \textbf{27.02}             \\
    \WIEMER     & 28.35                                            & 35.72                      \\
    \bottomrule
  \end{tabular}
\end{table}

With this work, we show and evaluate the reconstruction of the global magnetic perturbation field using spherical harmonics with LASSO regularization to promote sparsity in the coefficients. We show that it is possible to reconstruct from sparse measurements like the ones provided by \SUPERMAG stations, and we evaluated on \MHD data to show that the reconstruction is similar to the global field modeled by \MHD, suggesting that the results of such models can be compressed in spherical harmonics space by compressed sensing techniques. Additionally, we show that by using a Deep Neural Network to forecast in the spherical harmonics space, we can improve over existing state of the art (\WIEMER) by \SUPERMAGIMRPOVEMENT\; on \SUPERMAG and \MHDIMRPOVEMENT\; on \MHD dataset (summarized in table~\ref{performance}).

\section*{Broader Impact}

Geomagnetic storms drive a spectrum of potentially catastrophic disruptions to our technologically-dependent society, among the most threatening being critical disturbances to the electrical grid in the form of \textit{geomagnetically induced currents} (\code{GICs}). Due to their proprietary nature, publicly available \code{GIC} data are limited. However, a cohort study of insurance claims of electrical equipment provides evidence that space weather poses a continuous threat to electrical distribution grids via geomagnetic storms and \code{GICs}~\citep{Schrijver:2014, eastwood2018quantifying}. \code{GICs} also pose threats to oil pipelines, railways and telecommunication systems. In the case of extreme, but historically probable geomagnetic storms, the economic impact due to prolonged power outages can exceed billions of dollars per day ~\citep{Oughton:2017}. For this reason, there is urgency among public and industry stakeholders to improve monitoring and forecasting of space weather impacts like geomagnetic storms and \code{GICs}. With this work, we progress towards better and faster forecasting models which will help us shield our infrastructure from solar-wind related hazards.

\begin{ack}
This project was conducted during the 2020 NASA Frontier Development Lab (FDL) program, a public-private partnership between NASA, the SETI Institute, and commercial partners. We wish to thank, in particular, NASA, Google Cloud, NVIDIA, Intel and SRI, for supporting this project. Finally, we gratefully acknowledge the SuperMAG collaborators.
\end{ack}

\bibliography{geoeffectiveness}
\bibliographystyle{abbrvnat}

\appendix

\newpage

\section{Reconstruction experiment sweep parameter}

\begin{table}[!h]
  \caption{Sweep parameter description for reconstruction}
  \label{reconstruction_sweep}
  \centering
  \begin{tabular}{ll}
    \toprule
    Sweep parameter  & Description \\
    \midrule
    Date range       & 2015-06-23 to 2015-06-24\\
    Cadence          & 10 min \\
    Max modes sweep  & [5,50], steps of 5\\
    $\alpha$ sweep   & \{2,1,0.1,1e-3,1e-5\} \\
    Max L1 metric    & $\max_{\rm stations}\max_{\rm time}|| y-\tilde{y}||_1 $\\
    Max R2 score     & $\max_{\rm time} \left(1-  \sum_{\rm stations}|y-\tilde{y}|_1/\mathbf{\rm variance}(y)\right) $\\
    \bottomrule
  \end{tabular}
\end{table}

Note that we have used a cadence of 10 min for quick computation of the metrics over the dataset -- however, since the spherical harmonic generation is done at each time step, it can be performed at whatever cadence necessary.

\section{Solar Wind Data - \OMNI Dataset}

Here we describe the features we used from \OMNI Dataset.

\begin{table}[!h]
  \caption{Description of \OMNI Dataset features}
  \label{omni_features_description}
  \centering
  \begin{tabular}{ll}
    \toprule
    Feature  & Description \\
    \midrule
    \textbf{$B_T$:}       & Magnetic field magnitude ($\sqrt{(B_x^2+B_y^2)}$) \\
    \textbf{$V_{SW}$:}          & Solar Wind velocity magnitude ($\sqrt{(V_x^2+V_y^2+V_z^2)}$) \\
    \textbf{$T$:}  & Temperature of the solar wind\\
    \textbf{$\theta_c$:}   & Clock angle of the interplanetary magnetic field (IMF) \\
    \textbf{$F_{10.7}$:}    & F10.7 measures the noise level generated by the sun at a wavelength of 10.7 cm. \\
    \bottomrule
  \end{tabular}
\end{table}






\section{Geoeffectieness Indices}

\textbf{Dst:} (Disturbance Storm Time Index) It is the measure of geomagnetic activity derived from near equator ground magnetic stations providing information about the strength of ring current.

\textbf{Kp:} Global geomagnetic index that is based on 3 hour measurements of mid-latitude ground magnetic stations around the world.

\textbf{AE:} (Auroral Electrojet Index) It is the measure of auroral activity determined based on ground magnetic stations around aurora zone.

\section{\WIEMER}


\WIEMER baseline model is designed to forecast each of the three magnetic vector components. It uses solar wind data as input and it outputs the spherical harmonics coefficients, which then can be used to extract the forecasting values per CGM latitude and MLT pair.

A feature vector from the solar wind data is derived

\begin{align}
    F =
        \begin{pmatrix}
          \begin{bmatrix}
           1 \\
           B_T \\
           V_{SW} \\
           t \\
           \sqrt(F_{10.7}) \\
           B_T\cos(\theta_c) \\
           V_{SW}\cos(\theta_c) \\
           t \\
           \cos(\theta_c) \\
           \sqrt(F_{10.7}) \cos(\theta_c) \\
           B_T\sin(\theta_c) \\
           V_{SW}\sin(\theta_c) \\
           t \sin(\theta_c) \\
           \sqrt(F_{10.7})\sin(\theta_c) \\
           B_T \cos(2\theta_c) \\
           V_{SW}\cos(2\theta_c) \\
           B_T \sin(2\theta_c) \\
           V_{SW}\sin(2\theta_c)
          \end{bmatrix}
    \end{pmatrix},
  \end{align}

where $B_T$ represents the magnitude of the magnetic field, $V_{SW}$ the solar wind velocity, $\sqrt(F_{10.7})$ the square root of the $F10.7$ feature (a measure of solar radiation), $t$ the dipole axis angle in radians and $\theta_c$ the clock angle.

\WIEMER is trained on data obtained from \SUPERMAG stations for the year 2013 with the loss $\text{MSE}=\text{Avg}((y-Ba)^2)$, where $a$ are the spherical harmonics coefficients, computed as $a^m_n=(
\underbrace{g^m_n}_{\text{Real Part}}, \underbrace{h^m_n}_{\text{Imaginary Part}})$

The coefficients are computed as $g^m_n = G^m_n F$ and $h^m_n = H^m_n F$, where $G^m_n$ and $H^m_n$ are the weights to learn.

The model is trained on 25 minute long average of solar wind data, with a lag of 20 minutes and as target the magnetic field perturbation of 5 minutes long averages, as measured by \SUPERMAG stations.

\section{Reproducibility details of forecasting experiments}

The model architecture used for the forecasting experiment is presented below. The model was trained using Adam optimizer with learning rate \texttt{lr=1e-04} and Mean Squared Error as a loss.

\begin{minted}{python}
class GeoeffectiveNet(nn.Module):
    def __init__(self,
                 past_omni_length,
                 future_length,
                 omni_features,
                 supermag_features,
                 nmax,
                 targets):
        super(GeoeffectiveNet, self).__init__()

        self.omni_past_encoder = nn.GRU(25,
                                        16,
                                        num_layers=1,
                                        bidirectional=False,
                                        batch_first=True,
                                        dropout=0.5)

        n_coeffs = 0
        for n in range(nmax+1):
            for m in range(0, n+1):
                n_coeffs += 1
        n_coeffs *= 2

        self.encoder_mlp = nn.Sequential(
            nn.Linear(16, 16),
            nn.ELU(inplace=True),
            nn.Dropout(p=0.5),
            nn.Linear(16, n_coeffs, bias=False) # 882
        )

        self.omni_features = omni_features
        self.supermag_features = supermag_features

        self.targets = targets
        self.future_length = future_length

    def forward(self,
                past_omni,
                past_supermag,
                future_supermag,
                dates,
                future_dates,
                **kargs):

        past_omni = NamedAccess(past_omni, self.omni_features)

        features = []
        # add the wiemer2013 features
        bt = (past_omni['by']**2 + past_omni['bz']**2)**.5
        v = (past_omni['vx']**2 + past_omni['vy']**2 + past_omni['vz']**2)**.5

        features.append(past_omni['bx'])
        features.append(past_omni['by'])
        features.append(past_omni['bz'])
        features.append(bt)
        features.append(v)
        features.append(past_omni['dipole'])
        features.append(torch.sqrt(past_omni['f107']))

        features.append(bt*torch.cos(past_omni['clock_angle']))
        features.append(v*torch.cos(past_omni['clock_angle']))
        features.append(past_omni['dipole']*torch.cos(past_omni['clock_angle']))
        features.append(torch.sqrt(past_omni['f107'])*torch.cos(past_omni['clock_angle']))

        features.append(bt*torch.sin(past_omni['clock_angle']))
        features.append(v*torch.sin(past_omni['clock_angle']))
        features.append(past_omni['dipole']*torch.sin(past_omni['clock_angle']))
        features.append(torch.sqrt(past_omni['f107'])*torch.sin(past_omni['clock_angle']))

        features.append(bt*torch.cos(2*past_omni['clock_angle']))
        features.append(v*torch.cos(2*past_omni['clock_angle']))
        features.append(past_omni['dipole']*torch.cos(2*past_omni['clock_angle']))
        features.append(torch.sqrt(past_omni['f107'])*torch.cos(2*past_omni['clock_angle']))

        features.append(bt*torch.sin(2*past_omni['clock_angle']))
        features.append(v*torch.sin(2*past_omni['clock_angle']))
        features.append(past_omni['dipole']*torch.sin(2*past_omni['clock_angle']))
        features.append(torch.sqrt(past_omni['f107'])*torch.sin(2*past_omni['clock_angle']))

        features.append(past_omni['clock_angle'])
        features.append(past_omni['temperature'])

        features = torch.stack(features, -1)

        encoded = self.omni_past_encoder(features)[1][0]
        coeffs = self.encoder_mlp(encoded)

        predictions = torch.einsum('bij,bj->bi', future_supermag.squeeze(1), coeffs)

        return coeffs, predictions, None
\end{minted}

For the spherical harmonics decomposition we used \texttt{scipy\.special\.sph\_harm} function.

\begin{minted}{python}
  def basis_matrix(nmax, theta, phi):
    from scipy.special import sph_harm
    assert(len(theta) == len(phi))
    basis = []
    for n in range(nmax+1):
        for m in range(-n,n+1):
            y_mn = sph_harm(m, n, theta, phi)
            basis.append(y_mn.real.ravel())
            basis.append(y_mn.imag.ravel())
    basis = np.array(basis)
    return  basis
            .reshape(-1, theta.shape[0], theta.shape[1])
            .swapaxes(0, 1).swapaxes(2, 1)
\end{minted}

\section{International Geomagnetic Reference Field}

IGRF( International Geomagnetic Reference Field): Empirical measurements of the Earth magnetic field representing the main field without external sources.





\end{document}